\documentclass[aip,amsmath,amssymb,reprint]{revtex4-1}
\draft
\usepackage{graphicx}  
\usepackage{dcolumn}
\usepackage{subcaption}
\usepackage{amsmath}
\usepackage{makecell}
\usepackage{bm}        
\usepackage{amssymb}   
\usepackage[dvipsnames]{xcolor}
 
\usepackage[utf8]{inputenc}
\usepackage[T1]{fontenc}
\usepackage{mathptmx}
\usepackage{ragged2e} 
\usepackage{float}
\usepackage{color}

\begin{document}

\title{Characterizing the cage state of glassy systems and its sensitivity to frozen boundaries}

\author{Rinske M. Alkemade}
\affiliation{Soft Condensed Matter, Debye Institute of Nanomaterials Science, Utrecht University, Utrecht, Netherlands }
\author{Frank Smallenburg}
\affiliation{
Universit\'e Paris-Saclay, CNRS, Laboratoire de Physique des Solides, 91405 Orsay, France
}
\author{Laura Filion}
\affiliation{Soft Condensed Matter, Debye Institute of Nanomaterials Science, Utrecht University, Utrecht, Netherlands }
\begin{abstract}
Understanding the role that structure plays in the dynamical arrest observed in glassy systems remains an open challenge. Over the last decade, machine learning (ML) strategies have emerged as an important tool for probing this structure-dynamics relationship, particularly for predicting heterogeneous glassy dynamics from local structure.
A recent advancement is the introduction of the \textit{cage state}, a structural quantity that captures the average positions of particles while rearrangements are forbidden. During the caging regime, linear models trained on the cage state have been shown to outperform more complex ML methods trained on initial configurations only.
In this paper, we explore the properties associated with the cage state in more detail to better understand why it serves as such an effective predictor for the dynamics. Specifically, we examine how the cage state in a binary hard-sphere mixture is influenced by both packing fraction and boundary conditions. Our results reveal that, as the system approaches the glassy regime, the cage state becomes increasingly influenced by long-range structural effects. This influence is evident both in its predictive power for particle dynamics and in the internal structure of the cage state, suggesting that the CS might be associated with some form of an amorphous growing structural length scale. 
\end{abstract}

\maketitle
\section{Introduction}
For decades, the intriguing relation between structure and dynamics in glassy systems has been the topic of ongoing debate\cite{cavagna2009supercooled,berthier2011theoretical,royall2015role,tanaka2019revealing}. Whereas the structure remains seemingly unchanged when quenching the system, the dynamics are observed to slow down by orders of magnitude and become heterogeneous \cite{ediger1996supercooled}. Recently, one of the ways forward in probing this apparent discrepancy has been the use of machine learning (ML) strategies. Specifically, these ML algorithms try to predict the dynamical heterogeneity observed in glasses based on local structure. In recent years, there has been a rapid expansion in the variety of machine learning algorithms used to predict glassy dynamics\cite{Yang-2021, ciarella2023dynamics,ronhovde2011detecting,cubuk2015identifying, schoenholz2016structural,bapst2020unveiling,
boattini2020autonomously,paret2020assessing,boattini2021averaging,jung2023predicting, pezzicoli2022se,coslovich2022dimensionality, shiba2023botan,liu2024classification}, ranging from simple linear regression algorithms\cite{boattini2021averaging} to complex graph neural networks\cite{shiba2023botan}. In parallel, new and physically inspired structural parameters have been developed that better capture the local structure\cite{jung2023predicting,alkemade2023improving, sharma2024selecting,jiang2025enhancing}.

Interestingly, one observation that arose from these ML studies was that a simple linear regression model can make highly accurate predictions when the information provided on the local structure includes data on the so-called cage state \cite{alkemade2023improving, jung2025roadmap}. The cage state (CS), which was introduced in Ref. \onlinecite{alkemade2023improving}, is 
defined as the average positions of particles in their cage when no major rearrangements of the system are permitted. A key quantity associated with the CS is the absolute distance for each particle $i$, between its position in the initial configuration and in the CS $\Delta r_i^\mathrm{cage} = |\mathbf{r}_i^\mathrm{cage} - \mathbf{r}_i^\mathrm{init}|$. Including information on this distance significantly improves the dynamical predictions during the caging regime made by a linear regression model. Notably, during the caging regime the linear regression model trained on the CS even outperformed other more complex models that were trained on the initial structure only\cite{jung2025roadmap}. In addition, in a recently published paper\cite{jiang2025enhancing}, it was shown that including not only the magnitude of $\Delta \mathbf{r}_i^\mathrm{cage}$, but also its orientation in the set of structural parameters improved dynamical predictions made by a graph neural network. These results raise the question what structural features the CS exactly captures -- and why these features are difficult to extract from the initial configuration alone.
The goal of this paper will therefore be to explore the properties of the CS in more detail.

To this end, we study how the CS evolves in a  binary hard-sphere mixture as the system becomes more glassy. To analyze the obtained CSs, we study them both in terms of their correlation with the dynamical evolution of the system, as well as by considering their intrinsic structural features. In addition, we examine at what distance particles influence each other's cages by measuring the influence of boundary conditions.  To this end, we adopt a methodology inspired by point-to-set correlation measurements\cite{franz2007analytic,biroli2008thermodynamic}, in which particles outside a spherical cavity are frozen while the cage state is measured inside the cavity. 
We find that as the system becomes more glassy the CS becomes better defined and correlates stronger with the dynamics, a trend already observed by Jiang \textit{et al}\cite{jiang2025enhancing}. Moreover, we find an enhanced influence of boundary conditions with glassiness, suggesting that the CS might be associated with some form of an amorphous growing structural length scale. 

\begin{figure*}
    \centering
    \includegraphics[width=0.65\textwidth]{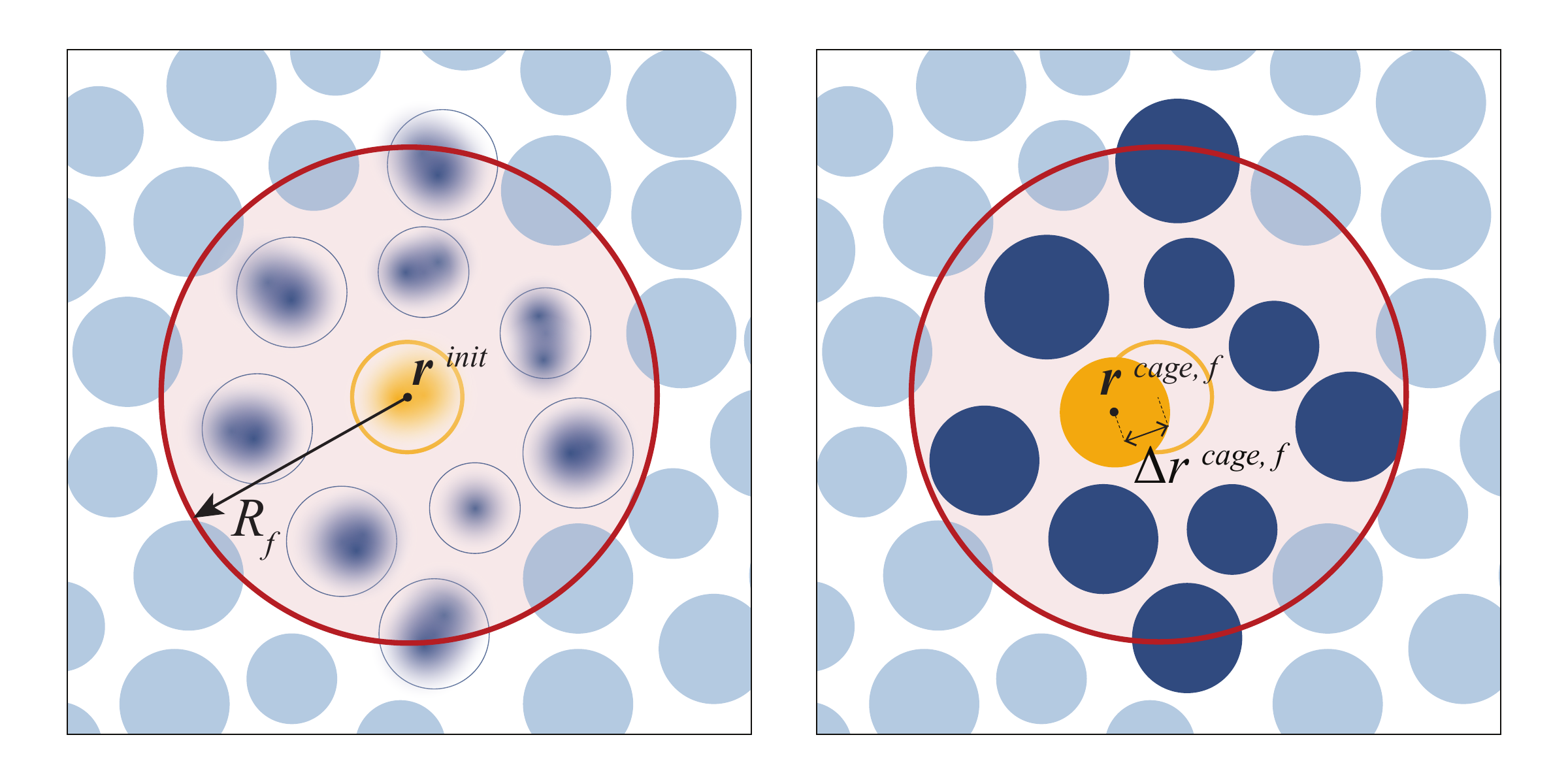}
    \caption{ Cartoon of the method used to measure the cage state as a function of freezing radius in a two-dimensional system. On the left, the cage state of the yellow particle is measured while allowing only particles within the red disk with radius $R_f$ to move. Particle motion is constrained using the spherical restriction method, meaning that particle $i$ is confined to the sphere with radius $\sigma_i/2$ centered around its initial position $\textbf{r}^\text{init}$. On the right, the cage state position $\textbf{r}^\text{cage, f}$ is displayed, along with the displacement vector, $\Delta r^\text{cage, f}$, between the initial position and the cage state.}
    \label{fig:Methodcagestate}
\end{figure*}

\section{Methods}
\subsection{Glass simulations}
As the glass-former model we use a binary hard-sphere system consisting of particle types $A$ and $B$ with respective particle diameters $\sigma_A$ and $\sigma_B$, where $\sigma_B/\sigma_A = 0.85$ and where the composition is given by $N_A/(N_A + N_B) = 0.3$. Note that this same mixture has been previously studied in several other works \cite{marin2020tetrahedrality, boattini2021averaging, boattini2020autonomously, alkemade2022comparing, alkemade2023improving}. We consider packing fractions in the range $\eta \in [0.53, 0.58]$ for systems of $N = N_A + N_B = 100{,}000$ particles. Additionally, for $\eta=0.58$ we also consider a smaller system consisting of $N=2000$ particles.

\subsection{Dynamic propensity}
In order to study the relation between the CS and the dynamical evolution of the systems, we use the well-known dynamical propensity as a measure of the dynamic heterogeneity of the system\cite{widmer2004reproducible, widmer2007study}. This measure, that captures the average mobility of particles, is obtained by sampling the isoconfigurational ensemble. To measure the dynamic propensity, we track the dynamical evolution of the system multiple times, each time starting from the same initial configuration, but with different initial velocities that are drawn randomly from a Maxwell-Boltzmann distribution at the desired temperature. Afterwards, we obtain the dynamic propensity $\Delta r_i(t)$ for particle $i$, by averaging the absolute displacement over the different runs as a function of time
\begin{equation}
    \Delta r_i(t) = \left\langle |\mathbf{r}_i(t) - \mathbf{r}_i(0)| \right\rangle_\mathrm{iso},\label{eq:propensity}
\end{equation}
where $\langle \dots\rangle_\mathrm{iso}$ indicates that we sample the isoconfigurational ensemble. Here, we average over 50 independent simulations.

To simulate the system we use event-driven molecular dynamic (EDMD) simulations \cite{Rapaport2009, smallenburg2022efficient} in the microcanonical ensemble where the energy $E$, the volume $V$ and the number of particles $N$ is fixed. To obtain equilibrated initial configurations, we first place $N$ particles with a reduced size in the box. These particles are then slowly grown over time, until the desired packing fraction is obtained. The system is then equilibrated in the canonical ensemble with fixed $N$, $V$ and $T$ (temperature) for at least $10\text{ }\tau_\alpha$, with $\tau_\alpha$ the relaxation time of the system, where time is measured in units of $\tau = \sqrt{m\sigma_A^2/k_BT}$, with $k_B$ Boltzmann's constant and $T$ the temperature (note that for $\eta = 0.58$ the relaxation time is given by\cite{boattini2020autonomously} $\tau_\alpha\approx10^4\tau$).

\subsection{Cage state}
To obtain the CS we follow the procedure outlined in Ref. \onlinecite{alkemade2023improving}. 
Given the initial configuration $\{\mathbf{r}_0^N\}$ the cage center of particle $i$, $\langle \mathbf{r}^\text{cage}_i \rangle$ is defined as the following restricted ensemble average

\begin{align}
    \langle \mathbf{r}^\text{cage}_i \rangle = \frac{\int d\mathbf{r}^N\mathbf{r}_i\exp^{-\beta \phi(\mathbf{r}^N)} g\left(\{\mathbf{r}^N\} ,\{\mathbf{r}_0^N\}\right)}{\int d\mathbf{r}^N\exp^{-\beta \phi (\mathbf{r}^N)}g\left(\{\mathbf{r}^N\} ,\{\mathbf{r}_0^N\}\right)},
    \label{eq:cagestatedef}
\end{align}
where $-\beta \phi(\mathbf{r}^N)$ is the potential energy of the system and where $g\left(\{\mathbf{r}^N\} ,\{\mathbf{r}_0^N\}\right)$ is a function that is zero when any particle moves too far away from its initial position. Note that the CS is a purely structural quantity, defined by a (constrained) ensemble average, and hence is not influenced by dynamics.   
Intuitively, the CS has some similarities to the inherent state, which is frequently used in glassy studies to identify an underlying structure independent of thermal fluctuations. The difference lies in the fact that, by definition, the inherent state locally minimizes the potential energy of the configurations. In contrast, the CS takes the average position of particles inside their cage and hence takes into account thermal fluctuations. As shown in Ref. \onlinecite{alkemade2023improving}, the CS outperforms the inherent state in terms of predicting future glassy dynamics. 

Note that there is some freedom in defining the function $g(\{\mathbf{r}^N\}, \{\mathbf{r}_0^N\})$ that constrains particles to their cages. Here, we adopt the same two restriction functions that we used in Ref. \onlinecite{alkemade2023improving}. 
In the first approach, each particle is confined to a spherical region with a fixed radius $r_c$ around its initial position  $\mathbf{r}_i^\mathrm{init}$, where here we choose $r_c$ to be equal to the particles radius. In the second approach, particles are restricted to an approximate definition of a Voronoi cell. For a particle $i$, this cell is defined by the set of points $\mathbf{R}$ for which
\begin{equation}
\frac{\left|\mathbf{R} - \mathbf{r}_i^\mathrm{init} \right|}{\sigma_i} < \frac{\left|\mathbf{R} - \mathbf{r}_j^\mathrm{init} \right|}{\sigma_j}  \forall j \in \mathcal{N}(i),
\end{equation}
where  $\mathcal{N}(i)$ denotes the nearest neighbours of particle \textit{i}, determined by the solid angle nearest neighbour (SANN) algorithm \cite{van2012parameter}.

\begin{figure}
    \centering
    \includegraphics[width=0.5\textwidth]{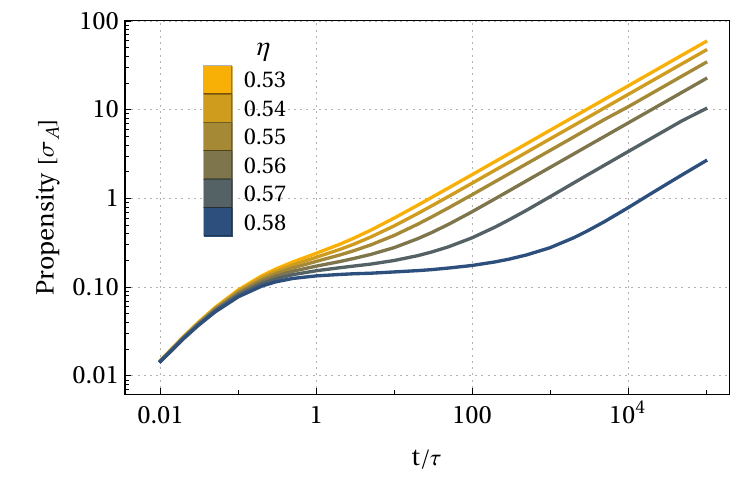}
    \caption{Propensity averaged over 30,000 $A$ particles plotted over time for a binary hard-sphere mixture at packing fractions ranging from $\eta = 0.53$ (yellow) to $\eta = 0.58$ (blue). The figure shows that with increasing packing fraction the caging regime, which is the regime where particle move around in a caged formed by their nearest neighbours, becomes more pronounced. }
    \label{fig:Propensity}
\end{figure}

\begin{figure*}[t!]
\begin{tabular}{lclc}
     &  && \\
     a) &  &b) &  \\[0cm]
     & \includegraphics[width=0.45\linewidth]{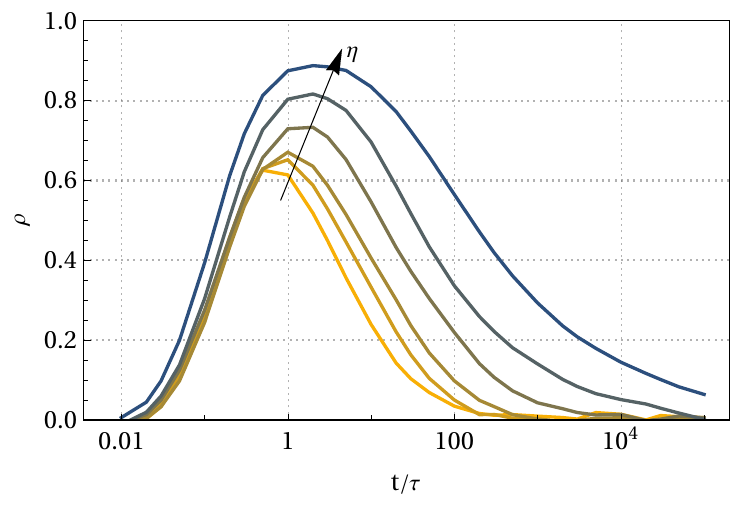} & &
      \includegraphics[width=0.45\linewidth]{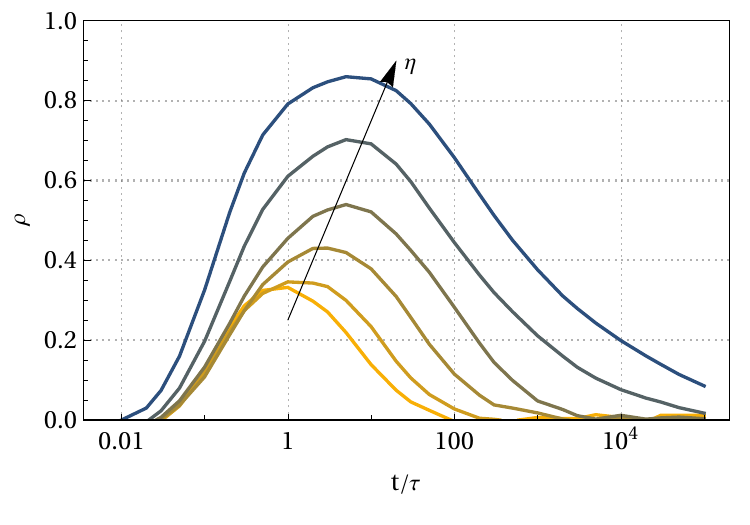} \\
\end{tabular}
    \caption[width=1\linewidth]{The Pearson correlation coefficient plotted as a function of time between the propensity and $\Delta r_i^\mathrm{cage}$ (which is the difference between the initial configuration and the cage state) for packing fractions in the range $\eta = 0.53$ (yellow) to $\eta = 0.58$ (blue) in steps of $\Delta \eta = 0.01$ for a) the spherical restriction method and b) the Voronoi restriction method. Correlations are based on $3\cdot10^4$ $A$-particles.}
    \label{fig:correlationdynamics} 
\end{figure*}

\subsection{Cage state in a partially frozen system}
In addition, we investigate the influence of boundary conditions on the CS by performing simulations where part of the system is held fixed. Inspired by the methodology used to measure point-to-set (PTS) functions, we freeze all particles located outside a spherical cavity of radius $R_f$ at their initial positions\cite{biroli2008thermodynamic}. 
In order to determine the CS under this restriction, we additionally restrict the movement of particles inside the cavity to their respective cages. To gain insight into the length scales over which particles influence each other's cage center, we focus solely on the effect of the frozen boundaries on the particle at the \textit{center} of the cavity. We therefore measure the cage center for each particle in an individual simulation,  where the unfrozen cavity is centered on the initial position of that particle (see Fig. \ref{fig:Methodcagestate}).
For each packing fraction, we measure the CS in the unfrozen system, as well as in cavities with a spherical radius in the range $R_f/\sigma_A\in[2,9]$. For each cavity size, the cage centers are measured for a subset of the $A$-particles, while for the unfrozen system the CS is measured for the entire system simultaneously. Simulations are performed using Monte-Carlo simulations where the system is equilibrated for $5\cdot 10^5$ cycles and measurements are performed for $10^6$ cycles.

\section{Results}

\subsection{Correlation between cage state and dynamic propensity}
We begin our study by exploring to what extent the structure of the CS captures information that is relevant for dynamical predictions. To get a feel for the glassiness of the systems we study in this paper, in Fig. \ref{fig:Propensity} we plot the dynamic propensity as a function of time for all investigated packing fractions. From this figure we see that the packing fractions we study range from fluid-like ($\eta \sim 0.53$) to glassy ($\eta \sim 0.58$) with a clear plateau indicating caging.

To examine the effect of density and cage restriction on the CS, in Fig. \ref{fig:correlationdynamics}, for six different packing fractions, we show the Pearson correlation coefficient as a function of time between the propensity and $\Delta r_i^\mathrm{cage}$ in an unfrozen system. Recall that $\Delta r_i^\mathrm{cage}$ is the distance between the initial position and the CS, given by $\Delta r_i^\mathrm{cage} = |\mathbf{r}_i^\mathrm{cage} - \mathbf{r}_i^\mathrm{init}|$.
In panel a) the cage state is obtained using the spherical restriction definition, while in panel b) we use the Voronoi cell restriction. These figures clearly show that as the packing fraction decreases, the correlation weakens and the peak shifts to the left. Note that this decrease in correlation as the system becomes less glassy has already been observed in Ref. \onlinecite{jiang2025enhancing} for the Kob-Anderson glass-model. The decrease is expected, since, as shown in Fig. \ref{fig:Propensity}, the time scale over which particles are trapped in their cages becomes shorter as we move to lower packing fractions. Hence, it is natural to expect that the CS plays less of a role in the dynamical evolution of the system.

Comparing the two panels in Fig. \ref{fig:correlationdynamics} we also see that with decreasing packing fractions the difference between the two cage restriction methods becomes more pronounced. For $\eta=0.58$, the two methods lead to almost similar correlations, whereas for lower packing fractions, the Voronoi-restriction is associated with significantly lower correlations. Apparently, at lower packing fractions the approximate Voronoi-cell mimics the cage less well than the spherical restriction. Note that by its definition, the space accessible to a particle under the Voronoi restriction is always strictly larger than under the spherical restriction. This difference in size increases with decreasing packing fraction, which likely explains why the Voronoi approach is less predictive at low packing fractions. For the remainder of this paper, we will therefore only show results obtained with the spherical-restriction method.

\begin{figure*}[htbp]
    \centering

    \begin{minipage}[t]{0.45\linewidth}
        \RaggedRight a)\\
        \includegraphics[width=\linewidth]{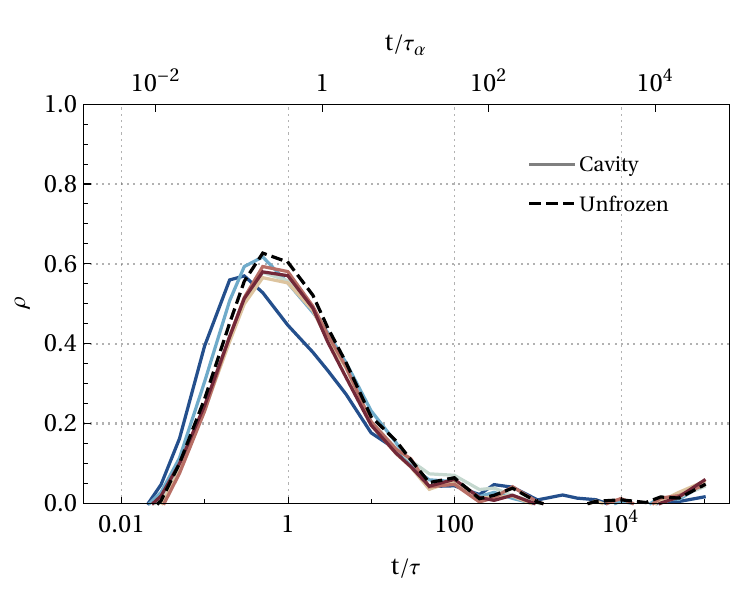}
    \end{minipage}
    \hspace{0.05\linewidth}
    \begin{minipage}[t]{0.45\linewidth}
        \RaggedRight b)\\
        \includegraphics[width=\linewidth]{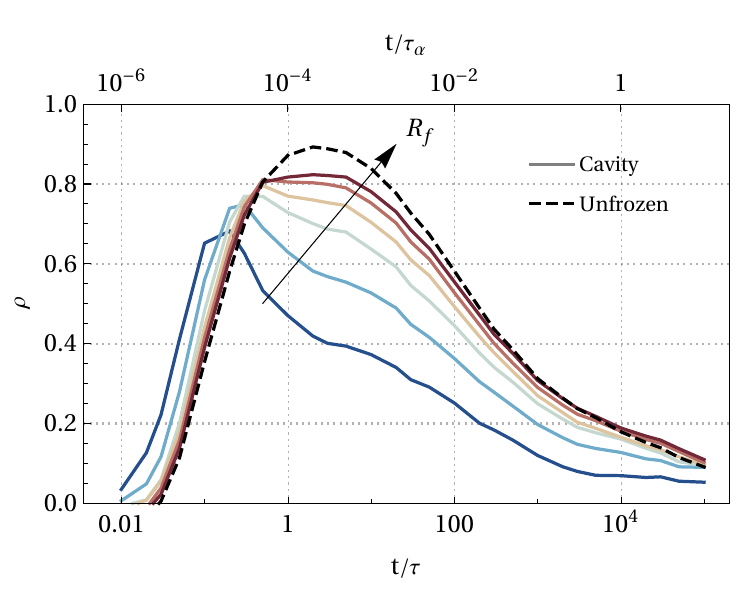}
    \end{minipage}

    \vspace{1em}

    \begin{minipage}[t]{0.45\linewidth}
        \RaggedRight c)\\
        \includegraphics[width=\linewidth]{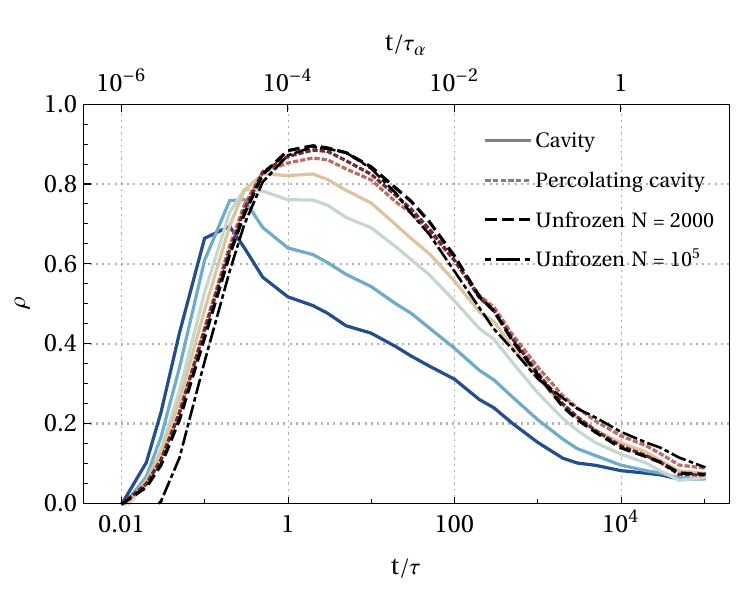}
        
    \end{minipage}
 \caption[width=1\linewidth]{The Pearson correlation coefficient between the propensity and $\Delta r_i^\mathrm{cage}$ as measured in a spherical cavity with different radii, plotted as a function of time. Time is expressed in units of $\tau$ (lower axis) and in units of the relaxation time $\tau_\alpha$ of the system (upper axis). Panel a) corresponds to a system at $\eta=0.53$ and with $N=10^5$ particles, panel b) to a system at $\eta=0.58$ and with $N=10^5$, and panel c) to a system at $\eta=0.58$ and with $N=2000$. The freezing radius  $R_f$ ranges from $R_f/\sigma_A = 2$ (blue) to $R_f/\sigma_A = 7$ (red) in steps of $\Delta R_f/\sigma_A = 1$. As a reference, the correlation associated with the unfrozen system is plotted as a black-dashed line. In panel c), additionally, the correlation associated with the large unfrozen system is also included (black dashed-dotted line). Colored dotted lines in panel c) are associated with cavity sizes that percolate the system. Note that all correlations are computed for $2000$ $A$-particles.}
  \label{fig:correlation5358} 
\end{figure*}   

Next, we explore the effect of boundary conditions on the correlation between the cage state and the propensity.  To this end, we measure the correlation between the propensity and $\Delta r_i^\mathrm{cage,f}(R_f)$, with $\Delta r_i^\mathrm{cage,f}(R_f)$ the distance between the initial configuration and the cage state obtained in a system where all particles outside of a spherical cavity with radius $R_f$ are frozen. In Fig. \ref{fig:correlation5358}a) and b), we show the correlations between the propensity and $\Delta r_i^\mathrm{cage,f}(R_f)$ for $\eta = 0.53$ and $\eta = 0.58$ respectively. In order to reduce computation times, we measure $\Delta r_i^\mathrm{cage,f}(R_f)$ for a sample of 2000 $A$-particles out of a total system size of $N=10^5$. In both panels, we also plot the correlation associated with $\Delta r_i^\mathrm{cage}$ measured for the same 2000 $A$-particles in a fully unfrozen system as a reference (dashed black line). 

For $\eta=0.53$, we observe that increasing $R_f \geq 4\sigma_A$ leads to correlations that are almost indistinguishable from the unfrozen system. This suggests that the structural characteristics of the cage state relevant for dynamical predictions are determined by particles within $R_f = 4\sigma_A$. In contrast, when looking at panel b), we observe that for $\eta=0.58$ the obtained correlation very strongly depends on the size of the cavity. Even for $R_f=7\sigma_A$ the correlation between propensity and $\Delta r_i^\mathrm{cage, f}(R_f)$ is significantly lower than in the unfrozen system. This result suggests that, as the system becomes more glassy, long-range boundary conditions have an enhanced influence on the CS, thereby weakening the correlation between CS and particle dynamics. This enhanced influence of the boundaries could potentially explain why ML strategies trained on initial configurations only have a hard time predicting the dynamics during the caging regime: estimating where a particle on average will be during the caging regime requires taking into account the relevant structure of the system over very large distances. Most existing ML approaches incorporate structural information through some form of averaging in order to reduce the parameter space\cite{bapst2020unveiling, boattini2021averaging, jung2023predicting}. Consequently, relevant large-scale structural information is almost inevitably lost in the process. In contrast, our CS computation has the advantage that the influence of particles far away is not averaged, thus preserving relevant structural information.

To study to what extent the system size influences the observed difference between the unfrozen CS and the CS measured in a cavity, we consider a much smaller system of 2000 particles at $\eta=0.58$.
In order to make a fair statistical comparison with the previous system size, we again consider a total of 2000 $A$-particles, obtained from four independent simulations.
In panel c) of Fig. \ref{fig:correlation5358} we show the correlation between $\Delta r_i^\mathrm{cage, f}(R_f)$ and the propensity for various freezing radii, as well as the unfrozen CS for this smaller system. As a reference, we also show the correlation for 2000 $A$-particles as obtained in the large unfrozen system at $\eta=0.58 $ consisting (dotted-dashed black line). From this figure we observe that there is no significant difference between the correlations associated with the small and large unfrozen system during the caging regime. Apparently, the amount of dynamical information embodied in the CS does not depend on the system size. This also means that large-scale density fluctuations, that could potentially play a role in the larger system, most likely do not influence the dynamics in the caging regime. 
Additionally, we see that, in contrast to the large system, for the small system the correlation for $R_f=7\sigma_A$ is almost indistinguishable from the unfrozen system. This is not unexpected since the smaller system has a box length of approximately $10.96 \sigma_A$. As a result, cavities with a radius larger $R_f \approx 5.48\sigma_A$ lead to a percolated system. Consequently, at $R_f=7\sigma_A$ almost all particles in the system can move. We speculate that the fact that even below $R_f \approx 5.48\sigma_A$ the correlations observed in the smaller system are higher than the corresponding correlations in the larger system is due to finite-size effects in the dynamic propensity (which is always measured in an unfrozen system). 

\begin{figure}
    \centering
    \includegraphics[width=0.49\textwidth]{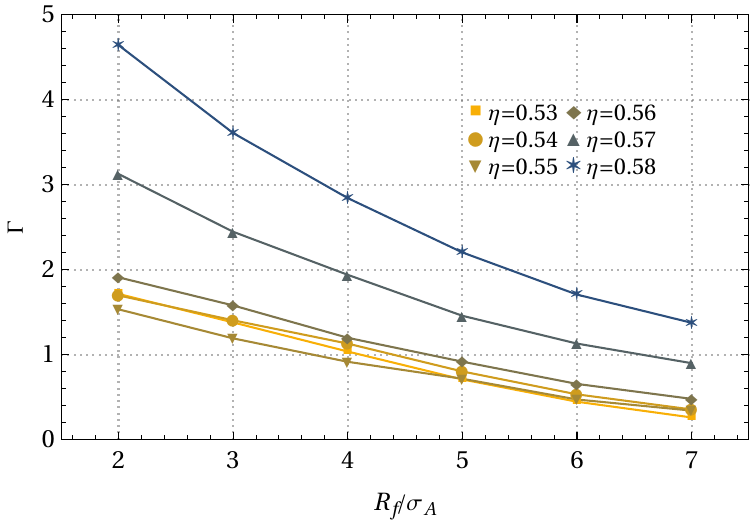}
    \caption{The figure displays the ratio $\Gamma$, defined in  Eq. \eqref{eq:rationodirection}, which quantifies the influence of frozen particles on the cage state. $\Gamma$ is plotted as a function of the freezing radius for six packing fractions in the range $\eta = 0.53$ to $\eta = 0.58$ for a system consisting of $N=10^5$. Averages are computed over $2000$ $A$-particles.  }
    \label{fig:Ratio}
\end{figure}

\begin{figure}
  
      \includegraphics[width=0.49\textwidth]{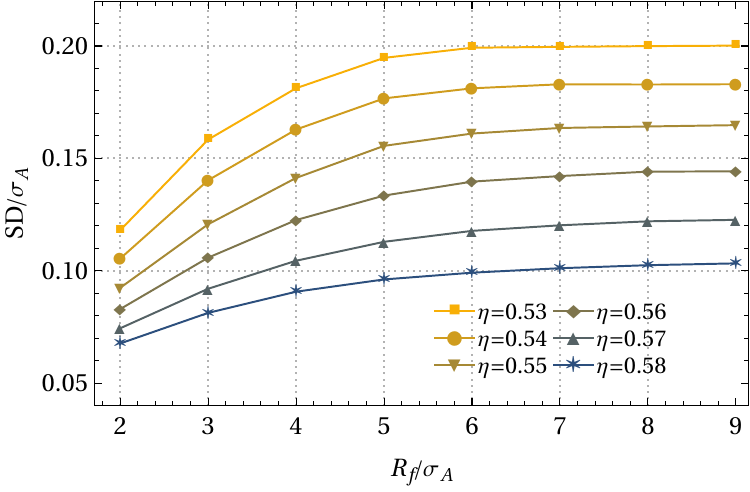}

    \caption[width=1\linewidth]{Average standard deviation of the non-averaged displacement vectors connecting particle's instantaneous positions to the associated cage center as a function of freezing radius (see Eq. \eqref{eq:SD}), plotted for various packing fractions in the range $\eta = 0.53$ to $\eta = 0.58$ for a system consisting of $N=10^5$. Averages are computed over $1000$ $A$-particles.}
    \label{fig:SDplot} 
\end{figure}
\subsection{Structure of the cage state}
Next, we examine the extent to which the measured CSs differ structurally across various cavity sizes. From this point on, we therefore no longer use the correlation with the dynamic propensity to evaluate the CS, but instead focus solely on its structural properties. When comparing the CS measured in a cavity to the CS of the unfrozen system, we note that the observed differences can be attributed to two main factors: boundary effects associated with the cavity, and thermodynamic noise arising from undersampling of the cage which is present in both the frozen and unfrozen systems. Since we are only interested in the influence of the boundaries, we want to isolate this last effect. In order to do so, we perform two independent simulations of the unfrozen system, generating two separate unfrozen CSs based on the same initial configuration, denoted by $\{\mathbf{r}^\text{cage,1}\}$, $\{\mathbf{r}^\text{cage,2}\}$, and compute the difference between those two CSs, $\{\mathbf{r}^\text{cage,1}- \mathbf{r}^\text{cage,2}\}$. This difference serves as a reference for the noise due to undersampling. We can then define a measure $\Gamma$, which captures the difference between the cavity CS and the unfrozen CS that can be attributed to boundary effects,
\begin{align}
    \Gamma = \frac{\sum_{i}|\mathbf{r}_{i}^\text{cage, f}(R_f)- \mathbf{r}^\text{cage,1}_{i}|}{\sum_{i}|\mathbf{r}^\text{cage,1}_{i}-\mathbf{r}^\text{cage,2}_{i}|} -1,
    \label{eq:rationodirection}
\end{align}
where the sum runs over all particles under consideration. When $\Gamma\approx 0$, the difference between the cavity cage center and the unfrozen cage centers is mainly due to thermodynamic noise, while a significant deviation from zero indicates boundary-induced differences. Note that we checked that when $R_f\rightarrow\infty$, $\Gamma$ is approximately zero.
In Fig. \ref{fig:Ratio} we plot $\Gamma$ as a function of freezing radius for six packing fractions. The results show that  at $R_f = 7\sigma_A$, boundary conditions influence the cage center within the cavity even at the lower packing fractions. However, it is clear that at higher packing fractions, the range over which the boundary conditions influence the cage state is significantly higher. Note that this enhanced sensitivity of the structure to boundary conditions as the system becomes more glassy has also been observed in point-to-set functions, see e.g. Ref. \onlinecite{biroli2008thermodynamic}, where it was attributed to a growing amorphous structural length scale. Although our results do not allow for a quantitative conclusion, they too suggest the presence of a growing amorphous structural length scale associated with the structure observed during the caging regime. 

Finally, we study the positions of particles within their cages in more detail. Specifically, during the cage state measurements, we track the non-averaged displacement vectors of individual particles, denoted as $\delta \mathbf{r}^{\text{cage}}$. These vectors connect particle's instantaneous positions to the associated cage centers. For a specific particle, the collection of these vectors thus captures the fluctuations of the particle around its cage center. To quantify these fluctuations we compute the average standard deviation of these vectors
\begin{equation}
    SD = \frac{1}{N_m}\sum_i\sqrt{\langle\delta \mathbf{r}_i^{\text{cage}}\cdot \delta \mathbf{r}_i^{\text{cage}}\rangle- \langle\delta \mathbf{r}_i^{\text{cage}}\rangle^2},
    \label{eq:SD}
\end{equation}
where the sum runs over $N_m$ particles. Note that by definition $\langle\delta \mathbf{r}_i^{\text{cage}}\rangle = \mathbf{0}$, such that the second term under the square root vanishes. Fig. \ref{fig:SDplot} shows the standard deviations as a function of freezing radius for the six packing fractions considered. As expected, the standard deviation is higher at lower packing fractions, reflecting the weaker caging of particles. Additionally, we find that the standard deviation increases monotonically with increasing freezing radius, indicating that the particles can fluctuate more as the cavity size increases. Notably, the standard deviation plateaus earlier for lower packing fractions. This suggests that, as the packing fraction increases, the freezing of particles increasingly affects not only the average positions of particles within their cages, but also the extent of their fluctuations.

\section{Conclusion}
In conclusion, we investigated how the CS evolves in a binary hard-sphere mixture as the system approaches the glassy regime, and how it is influenced by boundary conditions. To examine the latter, we measured the CS within an unfrozen spherical cavity, while pinning the surrounding particles to their initial positions. We characterized the CS both by measuring its correlation with dynamic propensity and by analyzing its structural features.

Overall, we found that the correlation between the CS and the dynamic propensity increases significantly with increasing packing fraction, indicating that the CS becomes better defined as the system becomes more glassy. In addition, we observed that with decreasing packing fraction, the choice of the restriction function for confining particles to their cages has more influence on the CS. At lower packing fractions, restricting particles to their approximate Voronoi cell led to significantly lower correlations with the propensity than the spherical restriction method. This difference can likely be attributed to the fact that, as the system becomes more liquid, a particle's Voronoi cell, which is strictly larger than the restricted spherical volume, mimics the cage less accurately. We expect this finding to be generalizable to other glass formers, meaning that the spherical restriction method is preferred also in those systems.

Furthermore, we found that the correlation between the propensity and the CS measured in an unfrozen cavity is strongly affected by the boundary conditions at large distances -- an effect that becomes more pronounced as the system approaches the glassy regime. The observed collective nature of the CS may help explain why machine learning techniques that are based solely on initial positions struggle to accurately predict particle dynamics during the caging regime: Our results suggest that, at high packing fractions, the structural information relevant to dynamical predictions is influenced by particles at large distances, implying that ML strategies must account for these non-local, collective effects in order to make good dynamical predictions.

In addition, we studied the structure of the CS by analyzing both the cage centers and the spatial spread of the particle around these centers across different packing fractions and freezing radii. Both quantities exhibited increasing sensitivity to the imposed boundary conditions with increasing packing fraction, which might indicate the presence of a growing amorphous structural length scale \cite{cavagna2007mosaic,franz2007analytic,biroli2008thermodynamic,bapst2020unveiling}.

These findings further establish the CS as a valuable structural quantity for capturing the collective nature of the caging regime and for probing the dynamics that govern particle dynamics during those timescales. Note that to date the CS has been used in ML studies. However, we expect it to be useful more generally for measuring quantities that depend on either the initial or inherent state, such as the local potential energy, local free volume, or Voronoi-based metrics \cite{ richard2020predicting, jung2023predicting}.

\section{Acknowledgments}
L.F. acknowledges funding from the Vidi research program with Project No. VI.VIDI.192.102, which is financed by the Dutch Research Council (NWO). 

\section*{Data Availability Statement}
All simulation codes needed to reproduce the data, as well as all relevant simulation data and notebooks to analyze the data and generate the figures is published as a data package in Ref.~\onlinecite{alkemade_2025_16947949}.

\section{References}
\addcontentsline{toc}{part}{Bibliography}
\markboth{\MakeUppercase{Bibliography}}{}
\bibliographystyle{abbrvunsrt2}
\bibliography{refs}

\begin{thebibliography}{10}

\bibitem{cavagna2009supercooled}
A.~Cavagna, {\em Supercooled liquids for pedestrians}, Phys. Rep. \textbf{476}, 51 (2009).

\bibitem{berthier2011theoretical}
L.~Berthier and G.~Biroli, {\em Theoretical perspective on the glass transition and amorphous materials}, Rev. Mod. Phys. \textbf{83}, 587 (2011).

\bibitem{royall2015role}
C.~P. Royall and S.~R. Williams, {\em The role of local structure in dynamical arrest}, Phys. Rep. \textbf{560}, 1 (2015).

\bibitem{tanaka2019revealing}
H.~Tanaka, H.~Tong, R.~Shi, and J.~Russo, {\em Revealing key structural features hidden in liquids and glasses}, Nat. Rev. Phys. \textbf{1}, 333 (2019).

\bibitem{ediger1996supercooled}
M.~D. Ediger, C.~A. Angell, and S.~R. Nagel, {\em Supercooled liquids and glasses}, J. Phys. Chem. \textbf{100}, 13200 (1996).

\bibitem{Yang-2021}
Z.~Y. Yang, D.~Wei, A.~Zaccone, and Y.~J. Wang, {\em Machine-learning integrated glassy defect from an intricate configurational-thermodynamic-dynamic space}, Phys. Rev. B \textbf{104}, 064108 (2021).

\bibitem{ciarella2023dynamics}
S.~Ciarella, M.~Chiappini, E.~Boattini, M.~Dijkstra, and L.~M. Janssen, {\em Dynamics of supercooled liquids from static averaged quantities using machine learning}, Mach. Learn.: Sci. Technol. \textbf{4}, 025010 (2023).

\bibitem{ronhovde2011detecting}
P.~Ronhovde, S.~Chakrabarty, D.~Hu, M.~Sahu, K.~Sahu, K.~Kelton, N.~Mauro, and Z.~Nussinov, {\em Detecting hidden spatial and spatio-temporal structures in glasses and complex physical systems by multiresolution network clustering}, Eur. Phys. J. E \textbf{34}, 105 (2011).

\bibitem{cubuk2015identifying}
E.~D. Cubuk, S.~S. Schoenholz, J.~M. Rieser, B.~D. Malone, J.~Rottler, D.~J. Durian, E.~Kaxiras, and A.~J. Liu, {\em Identifying structural flow defects in disordered solids using machine-learning methods}, Phys. Rev. Lett. \textbf{114}, 108001 (2015).

\bibitem{schoenholz2016structural}
S.~S. Schoenholz, E.~D. Cubuk, D.~M. Sussman, E.~Kaxiras, and A.~J. Liu, {\em A structural approach to relaxation in glassy liquids}, Nat. Phys. \textbf{12}, 469 (2016).

\bibitem{bapst2020unveiling}
V.~Bapst, T.~Keck, A.~Grabska-Barwi{\'n}ska, C.~Donner, E.~D. Cubuk, S.~S. Schoenholz, A.~Obika, A.~W. Nelson, T.~Back, D.~Hassabis, et~al., {\em Unveiling the predictive power of static structure in glassy systems}, Nat. Phys. \textbf{16}, 448 (2020).

\bibitem{boattini2020autonomously}
E.~Boattini, S.~Mar{\'\i}n-Aguilar, S.~Mitra, G.~Foffi, F.~Smallenburg, and L.~Filion, {\em Autonomously revealing hidden local structures in supercooled liquids}, Nat. Commun. \textbf{11}, 1 (2020).

\bibitem{paret2020assessing}
J.~Paret, R.~L. Jack, and D.~Coslovich, {\em Assessing the structural heterogeneity of supercooled liquids through community inference}, J. Chem. Phys. \textbf{152}, 144502 (2020).

\bibitem{boattini2021averaging}
E.~Boattini, F.~Smallenburg, and L.~Filion, {\em Averaging local structure to predict the dynamic propensity in supercooled liquids}, Phys. Rev. Lett. \textbf{127}, 088007 (2021).

\bibitem{jung2023predicting}
G.~Jung, G.~Biroli, and L.~Berthier, {\em Predicting dynamic heterogeneity in glass-forming liquids by physics-inspired machine learning}, Phys. Rev. Lett. \textbf{130}, 238202 (2023).

\bibitem{pezzicoli2022se}
F.~S. Pezzicoli, G.~Charpiat, and F.~P. Landes, {\em Rotation-equivariant graph neural networks for learning glassy liquids representations}, SciPost Phys. \textbf{16}, 136 (2024).

\bibitem{coslovich2022dimensionality}
D.~Coslovich, R.~L. Jack, and J.~Paret, {\em Dimensionality reduction of local structure in glassy binary mixtures}, J. Chem. Phys. \textbf{157}, 204503 (2022).

\bibitem{shiba2023botan}
H.~Shiba, M.~Hanai, T.~Suzumura, and T.~Shimokawabe, {\em Botan: Bond targeting network for prediction of slow glassy dynamics by machine learning relative motion}, J. Chem. Phys. \textbf{158}, 084503 (2023).

\bibitem{liu2024classification}
M.~Liu, N.~Oyama, T.~Kawasaki, and H.~Mizuno, {\em Classification of solid and liquid structures using a deep neural network unveils origin of dynamical heterogeneities in supercooled liquids}, J. Appl. Phys. \textbf{136}, 144702 (2024).

\bibitem{alkemade2023improving}
R.~M. Alkemade, F.~Smallenburg, and L.~Filion, {\em Improving the prediction of glassy dynamics by pinpointing the local cage}, J. Chem. Phys. \textbf{158}, 134512 (2023).

\bibitem{sharma2024selecting}
A.~Sharma, C.~Liu, and M.~Ozawa, {\em Selecting relevant structural features for glassy dynamics by information imbalance}, J. Chem. Phys. \textbf{161}, 184506 (2024).

\bibitem{jiang2025enhancing}
X.~Jiang, Z.~Tian, Y.~Hu, K.~Dong, W.~Hu, and Y.~Ai, {\em Enhancing glassy dynamics prediction by incorporating displacement from the initial to equilibrium state}, J. Phys. Chem. B \textbf{129}, 3053 (2025).

\bibitem{jung2025roadmap}
G.~Jung, R.~M. Alkemade, V.~Bapst, D.~Coslovich, L.~Filion, F.~P. Landes, A.~J. Liu, F.~S. Pezzicoli, H.~Shiba, G.~Volpe, et~al., {\em Roadmap on machine learning glassy dynamics}, Nat. Rev. Phys. \textbf{7}, 91 (2025).

\bibitem{franz2007analytic}
S.~Franz and A.~Montanari, {\em Analytic determination of dynamical and mosaic length scales in a {K}ac glass model}, J. Phys. A: Math. Theor. \textbf{40}, F251 (2007).

\bibitem{biroli2008thermodynamic}
G.~Biroli, J.~P. Bouchaud, A.~Cavagna, T.~S. Grigera, and P.~Verrocchio, {\em Thermodynamic signature of growing amorphous order in glass-forming liquids}, Nat. Phys. \textbf{4}, 771 (2008).

\bibitem{marin2020tetrahedrality}
S.~Mar{\'\i}n-Aguilar, H.~H. Wensink, G.~Foffi, and F.~Smallenburg, {\em Tetrahedrality dictates dynamics in hard sphere mixtures}, Phys. Rev. Lett. \textbf{124}, 208005 (2020).

\bibitem{alkemade2022comparing}
R.~M. Alkemade, E.~Boattini, L.~Filion, and F.~Smallenburg, {\em Comparing machine learning techniques for predicting glassy dynamics}, J. Chem. Phys. \textbf{156}, 204503 (2022).

\bibitem{widmer2004reproducible}
A.~Widmer-Cooper, P.~Harrowell, and H.~Fynewever, {\em How reproducible are dynamic heterogeneities in a supercooled liquid?}, Phys. Rev. Lett. \textbf{93}, 135701 (2004).

\bibitem{widmer2007study}
A.~Widmer-Cooper and P.~Harrowell, {\em On the study of collective dynamics in supercooled liquids through the statistics of the isoconfigurational ensemble}, J. Chem. Phys. \textbf{126}, 154503 (2007).

\bibitem{Rapaport2009}
D.~C. Rapaport, {\em {The Event-Driven Approach to {N}-Body Simulation}}, Prog. Theor. Phys. Supp. \textbf{178}, 5 (2009).

\bibitem{smallenburg2022efficient}
F.~Smallenburg, {\em Efficient event-driven simulations of hard spheres}, Eur. Phys. J. E \textbf{45}, 22 (2022).

\bibitem{van2012parameter}
J.~A. van Meel, L.~Filion, C.~Valeriani, and D.~Frenkel, {\em A parameter-free, solid-angle based, nearest-neighbor algorithm}, J. Chem. Phys. \textbf{136}, 234107 (2012).

\bibitem{cavagna2007mosaic}
A.~Cavagna, T.~S. Grigera, and P.~Verrocchio, {\em Mosaic multistate scenario versus one-state description of supercooled liquids}, Phys. Rev. Lett. \textbf{98}, 187801 (2007).

\bibitem{richard2020predicting}
D.~Richard, M.~Ozawa, S.~Patinet, E.~Stanifer, B.~Shang, S.~Ridout, B.~Xu, G.~Zhang, P.~Morse, J.-L. Barrat, et~al., {\em Predicting plasticity in disordered solids from structural indicators}, Phys. Rev. Mater. \textbf{4}, 113609 (2020).

\bibitem{alkemade_2025_16947949}
R.~Alkemade, F.~Smallenburg, and L.~Filion, {\em Data package: Characterizing the cage state of glassy systems and its sensitivity to frozen boundaries}, https://doi.org/10.5281/zenodo.16947949 (2025).

\end{thebibliography}

\end{document}